\documentclass[12pt]{spieman}  
\usepackage{amsmath,amsfonts,amssymb}
\usepackage{graphicx}
\usepackage{setspace}
\usepackage{tocloft}
\usepackage[round]{natbib} 
\usepackage{bbm}
\usepackage{booktabs}
\usepackage{siunitx}
\usepackage{float}
\usepackage{parskip}
\usepackage{caption}
\usepackage{comment}
\usepackage{color, colortbl}
\usepackage{xcolor}
\usepackage{titlesec}

\titleformat{\section}
  {\normalfont\large\bfseries}{\thesection.}{1em}{}
\titleformat{\subsection}
  {\normalfont\normalsize\bfseries}{\thesubsection.}{1em}{}
\titleformat{\subsubsection}
  {\normalfont\normalsize\itshape}{\thesubsubsection.}{1em}{}

\usepackage{tikz}
\usetikzlibrary{positioning}
\newdimen\nodeDist
\usepackage[figurename=Figure]{caption}
\usepackage{subcaption}
\usepackage{graphicx}
\DeclareCaptionFormat{custom}
{%
    \textbf{#1#2}\textit{\small #3}
}
\usepackage{etoolbox}
\apptocmd{\thebibliography}{\raggedright}{}{}

\title{Research Note: Bayesian Record Linkage with Application to Chinese Immigrants in Raleigh-Durham (ChIRDU) Study}

\author[a]{Eric A. Bai}
\author[a]{Madeleine Beckner}
\author[a]{Botao Ju}
\author[a]{Jerome P. Reiter}
\author[b]{Ted Mouw}
\author[c]{M. Giovanna Merli}
\affil[a]{Department of Statistical Science, Duke University, Durham, NC, USA}
\affil[b]{Department of Sociology and Carolina Population Center, University of Carolina at Chapel Hill, Chapel Hill, NC, USA}
\affil[c]{Sanford School of Public Policy and Duke Population Research Institute, Duke University, Durham, NC, USA}

\cftpagenumbersoff{figure}
\cftpagenumbersoff{table}

\begin{document} 
\maketitle

\begin{spacing}{2} 

\begin{abstract}
Many population surveys do not provide information on respondents' residential addresses, instead offering coarse geographies like zip code or higher aggregations.  However, fine resolution geography can be beneficial for characterizing neighborhoods, especially for relatively rare populations such as immigrants. One way to obtain such information is to link survey records to records in auxiliary databases that include residential addresses by matching on variables common to both files.  In this research note, we present an approach based on probabilistic record linkage that enables matching survey participants in the Chinese Immigrants in Raleigh-Durham (ChIRDU) Study to records from InfoUSA, an information provider of residential records. The two files use different Chinese name romanization practices, which we address through a novel and generalizable strategy for constructing records' pairwise comparison vectors for romanized names.  Using a fully Bayesian record linkage model, we characterize the geospatial distribution of Chinese immigrants in the Raleigh-Durham area.
\end{abstract}


\section*{Introduction}
\label{sect:intro}  
Over the last four decades, the heterogeneity of Asian immigration to the U.S.\ has increased in terms of origin and demographic profile \citep{ruiz,pewresearchcenter}. Similar to other new immigrant groups, Asian immigrants have become increasingly dispersed across suburban areas \citep{leejen, flippen} and new areas of destination \citep{flip,sakamoto}, moving away from concentrated ethnic neighborhoods in the largest cities of the traditional gateway areas of California, Texas, Illinois, and the Northeast. 
As a result of the increased spatial dispersion of immigrants, the foreign-born can represent small proportions of local populations  
\citep{parado}. This is the case of individuals of Chinese descent in North Carolina, who are part of a rapidly growing Asian population group but still account for less than 1\% of the area’s residents \citep{tippett_2018, unc-aac}.

Immigrants’ spatial distribution has important implications for understanding their social incorporation as measured by residential patterns of spatial assimilation. To describe this spatial distribution, it is beneficial to use finely-resolved geographic information on the immigrants' residences.  For example, 
population counts aggregated at the zip code or higher level  may  not be granular enough to describe the characteristics of neighborhoods where Chinese immigrants reside  (or other population groups, for that matter). 
Address-level information often is not collected in surveys, for example, because the modality of data collection does not require this collection.  In such settings, researchers may be able to obtain addresses by linking the study participants' data to  records from auxiliary data sources that include such information.  Typically, this matching cannot be done using direct identifiers like social security numbers, as they are not available or accessible on one or both of the files.  Instead, researchers need to match records on imprecise fields like names or demographic characteristics using  probabilistic record linkage techniques \citep{fellegi1969theory}.  Once linked, the records can enable a variety of analyses, for example, examining whether or not  people with certain characteristics tend to cluster geographically or the characteristics of neighborhoods where immigrants in new areas of destinations are dispersing.

In this research note, we use a probabilistic record linkage approach to match participants in a local, first-generation immigrant population representative survey, the Chinese Immigrants in Raleigh-Durham (ChIRDU) Study \citep{Merli}, to individual/household records from InfoUSA, a private-sector provider of residential data. Since participants in ChIRDU vary in their places of origin (e.g., Taiwan/mainland,  China/Hong Kong), their Chinese names may be romanized differently according to each region’s practices. The ChIRDU investigators standardized participants' names according to the official romanization system used in mainland China, \emph{pinyin}\footnote{\emph{Pinyin} is the standard, most common romanization scheme for Chinese characters used in mainland China and internationally. Beginning October 1, 2000,  American libraries began 
using {\em pinyin}, 
replacing the Wade-Giles system used by American libraries in the 20th century (\url{https://www.loc.gov/item/prn-00-141/}).}; however, it is not clear which standardization method, if any, is employed by InfoUSA.  Thus, if we match on names, whether by  requiring an exact match or by using some typical  metric that scores the similarity of the romanized names for record pairs, we are likely to miss some matches.  To account for this possibility, we propose and utilize a novel comparison variable for record pairs' names. In particular, this variable accounts for four levels of agreement on name: exact match, match on a 
set of alternative ways Chinese names may have been romanized, and two levels based on traditional string metrics that account for  differences in spelling.
We adopt a Bayesian probabilistic record linkage approach \citep{sadinle} that allows for propagation of uncertainty through multiple plausible linkage outputs. Using the posterior samples, we examine the geospatial distributions along several dimensions of the Chinese immigrants in the ChIRDU data.  We conclude by showcasing several ways to examine the performance of the record linkage procedure.

The linkage of ChIRDU participants' data to InfoUSA data was approved by the IRB at Duke University 
and involved multiple precautions to reduce risks to data subjects' confidentiality.  These are described in the supplementary material.

\section*{Overview of Record Linkage Model}

In this section, we review the setup for the  model of \citet{sadinle}, which is a Bayesian version of the probabilistic record linkage approach of \citet{fellegi1969theory}.  Details of the model are available in the supplementary material. We implement the model using the ``BRL'' package in R.


 Let $\mathcal{A}$ and $\mathcal{B}$ be two databases with $n_{A}$ and $n_{B}$ records, respectively.  Both $\mathcal{A}$ and $\mathcal{B}$ include $K$ variables in common, which we refer to as linking fields.  Let $\mathcal{A}= \{\mathcal{A}_i: i=1, \dots, n_A\}$, where  each record $\mathcal{A}_i = (\mathcal{A}_{i1}, \dots, \mathcal{A}_{iK})$ is measured on the $K$ variables.  Similarly, let $\mathcal{B}= \{\mathcal{B}_j: j=1, \dots, n_B\}$, where  each record $\mathcal{B}_j = (\mathcal{B}_{j1}, \dots, \mathcal{B}_{jK})$.   Our setting is a bipartite record linkage task, in which  
 each individual is represented at most once within each of $\mathcal{A}$ or $\mathcal{B}$, that is, no duplication exists within a database.  Both $\mathcal{A}$ and $\mathcal{B}$ may have other variables as well.  We do not concern ourselves with such variables for purposes of record linkage, although it is possible to utilize them to improve record linkage quality \citep[e.g., ][]{gutman, dalzellreiter, larsenlahiri, wortmanreiter}.



Intuitively, two records $\mathcal{A}_i$ and $\mathcal{B}_j$ that belong to the same individual should have similar values of most of the $K$ linking fields, whereas two records that belong to different individuals should not. 
We formalize this intuition by using comparison vectors.  
For every record $\mathcal{A}_i$ and $\mathcal{B}_j$, and every linking field $k=1, \dots, K$,  
we define $\gamma_{ij}^k$ to take on levels of agreement in $\{1, \dots, d_k\}$. Here,  smaller values of $\gamma_{ij}^k$ indicate closer agreement on field $k$, and larger values indicate the opposite. For example, suppose we use zip code as the $k$th linking field, and we define the comparison as identical zip codes or not.  We would define a binary  $\gamma_{ij}^k$, i.e., with $d_k=2$.  We let $\gamma^k_{ij} = 1$ when records $\mathcal{A}_i$ and $\mathcal{B}_j$ have identical values of zip code,
and $\gamma^k_{ij} = 2$ otherwise. 


For linking fields that involve strings, e.g., names or addresses, it is common to define the corresponding $\gamma_{ij}^k$ by using string similarity/distance metrics \citep{christian}. 
This a function that maps two vectors of characters to a non-negative number, with higher values indicating greater similarity between the vectors. We then discretize these scores into distinct threshold/cutoffs to facilitate record linkage modeling. 
Several well-established string similarity metrics utilize edit-based distance, which counts the number of edit operations (substitution, deletion, insertion) required to transform one string into another. The Jaro-Winkler metric, introduced by \citet{jaro} and improved by \citet{winkler}, gives heavier weight to the prefix similarities of strings and is more suitable on names where abbreviations/typos may occur at the beginning of the string. Another popular edit-distance metric is the Levenshtein distance, which calculates the minimum number of single-character edit operations required to transform one string into another \citep{levenshtein1966binary}. While many edit distance metrics have found success in summarizing words of western origin, they have limitations as measures of romanized strings with non-western origins (romanized Chinese names), as we illustrate below.


We now define some notation from the model of \citet{sadinle} useful for summarizing results of the Bayesian record linkage.  Suppose that we seek to match records in $\mathcal{A}$ (in our context, InfoUSA) to records in $\mathcal{B}$ (in our context, ChIRDU), and that $n_A>n_B$. We define the vector $\mathbf{Z} = (Z_1, \dots, Z_{n_B})$ such that, for each $\mathcal{B}_j$ where $j=1, \dots, n_B$,  
\begin{align*}
Z_j=
    \begin{cases}
        i, & \text{if $\mathcal{B}_j$ matches to $\mathcal{A}_i$;} \\
        n_A+j, & \text{if $\mathcal{B}_j$ does not have a match in $\mathcal{A}$.} 
    \end{cases}
\end{align*}
The $\mathbf{Z}$ fully defines the linkage structure for all $n_An_B$ record pairs. Of course, we do not know $\mathbf{Z}$ and must estimate it from the data in the comparison vectors. The BRL package provides an estimate of  the posterior distribution of $\mathbf{Z}$ using a Markov chain Monte Carlo (MCMC) sampling algorithm. As a byproduct of the MCMC sampling, the BRL package generates many, say $H$, plausible draws of the linked data files, $(\mathbf{Z}^{(1)}, \dots, \mathbf{Z}^{(H)})$. 

For each record in $\mathcal{B}_j$, we use the plausible draws to estimate the posterior probability that any $\mathcal{A}_i$ matches to any $\mathcal{B}_j$. Simply, this is the fraction of times among that $H$ plausibly linked files that the two records are matched.  We also can compute the fraction of times $Z_j=n_A+j$ to estimate the probability that $\mathcal{B}_j$ does not have a link in $\mathcal{A}.$  As a  point estimate of the linkage structure, we use the most frequently occurring draw of $\mathbf{Z}$ among the $H$ plausibly linked files.




\section*{ChIRDU and InfoUSA Linkage}
In this section, we summarize features of the ChIRDU and InfoUSA data sets and describe the comparison vectors we use in the Bayesian record linkage model.

\subsection*{Data Sets}
The primary data of interest were collected between March 2018 and January 2019 \citep{Merli}. The sample was recruited with Network Sampling with Memory \citep{mouw}, a link-tracing sampling design which recruits respondents from their social networks. The sample consists of 509 age 18+ respondents born in mainland China, Taiwan or Hong Kong who resided in the Raleigh-Durham (RDU) area at the time of the survey. Recruitment began with seven seed respondents known to the principal investigators of the project or to the field interviewers. 502 participants were recruited through an iterative referral process. Information on their socioeconomic characteristics, education, acculturation, zip codes, and a host of other demographic variables was collected from all 509 participants. The sample was shown to be representative of the Chinese foreign-born population of the Raleigh-Durham area when compared with the American Community Survey \citep{Merli}. 



InfoUSA (now Data Axle USA) is a marketing and sales company specializing in the provision of information on households and their characteristics (including, notably for our purposes, individuals' age, ethnicity, and household address coordinates) across the nation. The database is built and maintained using  records from  census statistics, billing statements, telephone directory listings and mail order buyers/magazine subscribers. According to their website \url{https://www.dataaxleusa.com/}, for 2018 InfoUSA has data on 309 million people and their respective households. InfoUSA exhibits lower coverage rates on households than those of the decennial population census; however, its coverage is comparable to other national sampling frames, and, in urban areas, provides strong coverage of addresses at the Basic Street Address level \citep{kennel2009content}.   

To obtain InfoUSA's Chinese population in the Raleigh-Durham area, we use the 2018 InfoUSA data and discard  all households with zip codes not among those reported by participants of ChIRDU, who reside in one of the three counties making up the Raleigh-Durham region of North Carolina: Durham, Wake and Orange.  We then  leverage the database's ethnicity code variable to include only individual households with at least one household member identifying as Chinese (``CN'' ethnicity code).  InfoUSA uses proprietary name evaluation algorithms to guess at each  individual's ethnicity based on ethno-linguistic and geocentric rules \citep{axle}. As such, to the extent the InfoUSA ethnicity imputations are inaccurate, our subset of InfoUSA data may exclude some Chinese immigrants who InfoUSA mislabeled as non-Asian.   
In all, the filtered InfoUSA data comprise 18,934 individuals who potentially could be links for the 509 ChIRDU participants.

\subsection*{Comparison Vectors}
\label{sec:Methodology}


The ChIRDU and InfoUSA data have four variables in common that we  use as linking fields: individual's first name, 
last name,
zip code, 
and age. 
We let $\gamma_{i,j}^1$ be the comparison for first name, $\gamma_{i,j}^2$ the comparison for  last name, $\gamma_{i,j}^3$ the comparison for zip code, and $\gamma_{i,j}^4$ the comparison for age.

\begin{figure}[t]
\centering
\scalebox{0.75}{
\begin{tikzpicture}[
    node/.style={%
      draw,
      rectangle,
    },
  ]

    \node [node] (A) {Exact Match?};
    \path (A) ++(-135:\nodeDist) node [node] (B) {$\gamma^k_{i,j}=1$};
    \path (A) ++(-45:\nodeDist) node [node] (C) {Exact Match on Wade-Giles?};
    \path (C) ++(-135:\nodeDist) node [node] (D) {$\gamma^k_{i,j}=2$};
    \path (C) ++(-45:\nodeDist) node [node] (E) {Jaro-Winkler $>0.9$};
    \path (E) ++(-135:\nodeDist) node [node] (F) {$\gamma^k_{i,j}=3$};
    \path (E) ++(-45:\nodeDist) node [node] (G) {$\gamma^k_{i,j}=4$};

    \draw (A) -- (B) node [left,pos=0.25] {yes}(A);
    \draw (A) -- (C) node [right,pos=0.25] {no}(A);
    \draw (C) -- (D) node [left,pos=0.25] {yes}(A);
    \draw (C) -- (E) node [right,pos=0.25] {no}(A);
    \draw (E) -- (F) node [left,pos=0.25] {yes}(A);
    \draw (E) -- (G) node [right,pos=0.25] {no}(A);
\end{tikzpicture}
}
\caption{Decision rule for constructing $\gamma_{ij}^k$ for name comparisons.  Here, $k=1$ for first names and $k=2$ for last names.} \label{figure:Decision Tree}
\end{figure}
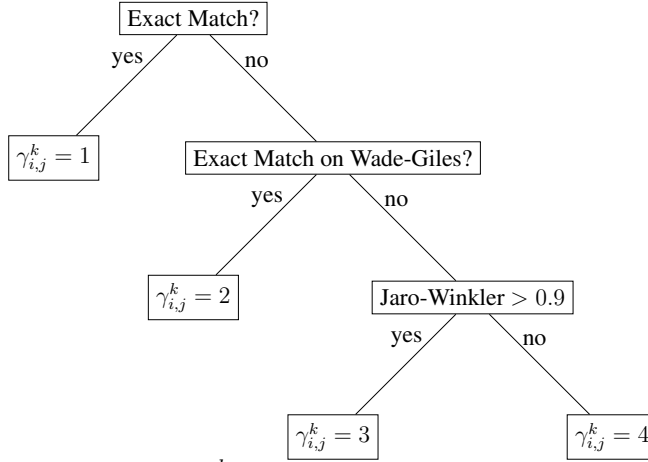

Record linkage tasks with Chinese individuals can be very challenging due to naming conventions, transliteration ambiguities, and different romanization systems, though novel pre-processing approaches have been introduced to address these challenges for historical (pre-\emph{pinyin}) Chinese names \citep{POSTEL2023101493}. Although all ChIRDU participants' names  were standardized to \emph{pinyin}, there is a need to account for possible different romanizations of Chinese names in InfoUSA. We therefore obtain the 
Wade-Giles equivalents \footnote{These are listed at \url{https://www.britannica.com/topic/Wade-Giles-romanization}.} of all {\em pinyin} characters. 
We then construct $\gamma_{ij}^1$ and $\gamma_{ij}^2$ to account for multiple levels of agreements in the first and last names. We set $\gamma^1_{ij} =1$ when the first names for record $\mathcal{B}_j$ in ChIRDU and record $\mathcal{A}_i$ in InfoUSA match exactly. When these first names do not match exactly, we check for an exact match between the Wade-Giles equivalent of the ChIRDU name and the original name in InfoUSA.  When these match, we set  $\gamma^1_{ij} =2$ indicating an exact match to the conversion. When neither of these two conditions hold,  we compare the two first names using the Jaro-Winker distance \citep{winkler}.  When this similarity score exceeds 0.9, indicating that the names generally agree except for a small number of letters, we set $\gamma^1_{ij} =3$; when not, we set $\gamma^1_{ij} =4$.  Thus, the entire process can be summarized as initially checking for exact matches on first names, if not then exact matches on converted names, if not then assessing the similarity of values on the first names, e.g., they differ because of some typographical error in one of the files.  Figure \ref{figure:Decision Tree} characterizes the decision rules for constructing the comparison vectors.  We use a similar procedure for last names.  

Table \ref{table:toy comp} illustrates determinations of several record pairs' $\gamma_{ij}^2$ on artificial data. In this example, we seek to link three records  $(\mathcal{B}_1, \mathcal{B}_2, \mathcal{B}_3)$ in one file potentially to four records $(\mathcal{A}_1, \mathcal{A}_2, \mathcal{A}_3, \mathcal{A}_4)$ in another file.  Record $\mathcal{B}_2$ has {\em pinyin} surname Cai with Wade-Giles equivalent of Tsai, whereas record $\mathcal{B}_1$ has surname Wang, which does not have a different Wade-Giles equivalent.  
 After applying the decision rules from Figure \ref{figure:Decision Tree}, we see that the names for $\mathcal{B}_1$ and $\mathcal{A}_1$  match exactly, whereas the names for $\mathcal{B}_1$ and the three other records in $\mathcal{A}$ are quite different and thus have low Jaro-Winkler scores.  The names for $\mathcal{B}_2$ and $\mathcal{A}_2$ are not identical, but they are Wade-Giles equivalents. Finally, records $\mathcal{B}_3$ and $\mathcal{A}_3$ are Wade-Giles equivalents, whereas $\mathcal{B}_3$ and $\mathcal{A}_4$ are similar enough that the Jaro-Winkler score exceeds 0.9. This final comparison  demonstrates a limitation of using only a Jaro-Winkler string-distance score for certain Chinese names. Zhen and Zheng in \emph{pinyin} are totally different surnames in Chinese, but they nevertheless achieve a high Jaro-Winkler score. 


In total, 364 (71\%) of the 509  ChIRDU participants have a Wade-Giles equivalent first or last name that differs from the {\em pinyin} equivalent. Among the approximately 9.6 million pairwise comparisons for names, around 10,000 matched on {\em pinyin} first name, 140,000 matched on {\em pinyin} last name, and only 441 matched on both {\em pinyin} first and last name. Approximately 16,000 comparisons of names in ChIRDU and InfoUSA matched exactly on the transformed (Wade-Giles) name, and approximately 113,000 comparisons have a Jaro-Winkler score exceeding 0.9. The vast majority of comparisons, approximately 8 million, fall below the Jaro-Winkler 0.9 threshold.

The comparison vectors for age and zip codes use binary comparisons. We set $\gamma^3_{ij}=1$ when the five-digit zip codes for $\mathcal{A}_i$ and $\mathcal{B}_j$ match exactly, and $\gamma^3_{ij}=2$ otherwise.  We set $\gamma^4_{ij}=1$ when the ages for $\mathcal{A}_i$ and $\mathcal{B}_j$ are both within the same age bin, and $\gamma^4_{ij}=2$ otherwise.  The bins include one for ages less than 25, then ten bins of five-year intervals (e.g., $25 \leq $ age $ <30$), and finally one bin for ages 75 or older. We use five-year age bins to allow for potential  errors of $\pm 2$ years in the 
birth year data in InfoUSA or ChIRDU.  We note that the youngest and oldest age bins in ChIRDU include low numbers of individuals.

\section*{Linkage Analysis}

We fit the Bayesian record linkage model described in the supplementary material using the BRL package.  We run the MCMC sampler for 10,000 iterations, discarding the first 1,000 samples as burn-in. Trace plots of model parameters do not suggest any lack of convergence.  

Using the $H=9000$ draws of  $\mathbf{Z}$, we estimate a posterior mode of 196 individuals in ChIRDU having a link in InfoUSA, with a 95\% posterior interval of $(182, 219)$ individuals.  Conversely, we estimate 313 individuals in ChIRDU do not have a link in InfoUSA, with a 95\% interval of $(290, 327)$.  The roughly 39\% match rate reflects the challenge of precisely distinguishing Chinese names with an absence of standardized conventions for romanization. Additionally, it is possible that some of the ChIRDU participants are not in the InfoUSA data (this is certainly the case vice versa).  Nonetheless, compared to only obtaining 64 individuals who match exactly on the four linking fields, the much higher match rate from the Bayesian probabilistic record linkage 
underscores the potential benefits of using the name comparison vectors and modeling approach.  

The posterior distribution  of $\mathbf{Z}$ facilitates investigation of the ChIRDU participants' geospatial distributions. 
We interpret each of the 9,000 draws of $\mathbf{Z}$ as a plausible set of linked data.  Using these, we can estimate the spatial distribution of the residences of the matched Chinese immigrants. In particular,  we can visualize these plausible linkages by plotting the locations for several randomly sampled vectors from $\mathbf{Z}^{(1)}, \dots, \mathbf{Z}^{(H)}$. 

\begin{figure}[t]
\centering
\includegraphics[width=0.85\textwidth]{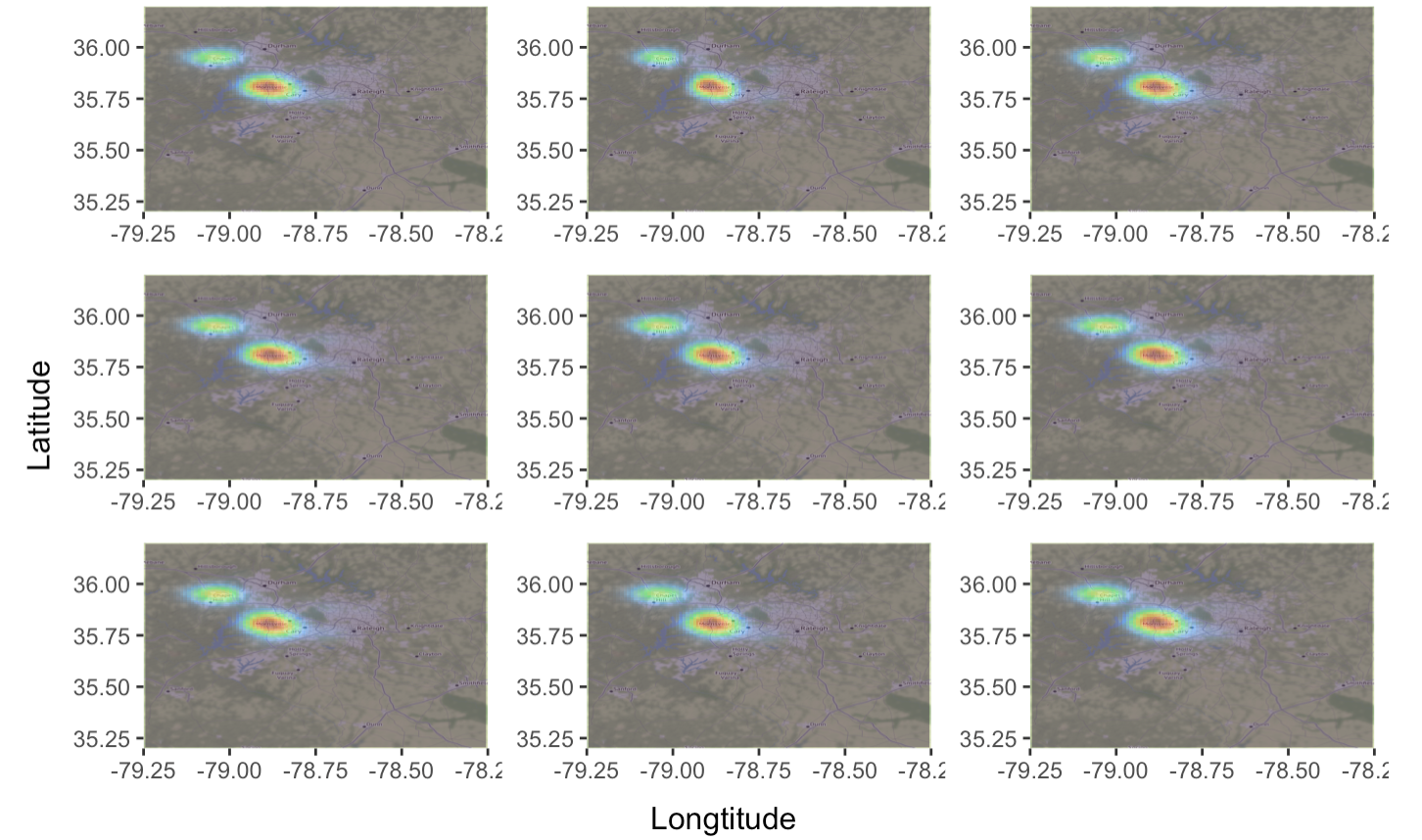}
\caption{Match density for nine random iterations of the MCMC sampler.}
\label{fig:posterior_densitymap}
\end{figure}

Figure \ref{fig:posterior_densitymap} displays density plots of the residence locations for the matches over nine random iterations of the MCMC sampler. We use density plots rather than individual point locations to  protect the confidentiality of ChIRDU respondents' locations as required in our approved IRB protocol for publication of  results.\footnote{Our pre-publication plots are at the address level and reveal finer details of the spatial distribution.}    
While the plots exhibit some differences reflecting the posterior uncertainty in the distributions, overall they are quite similar, suggesting that we can draw robust conclusions about the records that can be linked.  Specifically, the matches tend to occur in two general locations, corresponding roughly to the towns of Chapel Hill and Morrisville/Cary.  This distribution aligns with expectations, as these are areas well known for their diverse populations and a strong presence of immigrants born in China \citep{tippett_2018} employed in nearby universities,  health systems, and IT and pharmaceutical companies.

. 





\begin{figure}[t]
\noindent\begin{minipage}{0.48\textwidth}
     \centering
     \includegraphics[width = 8cm]{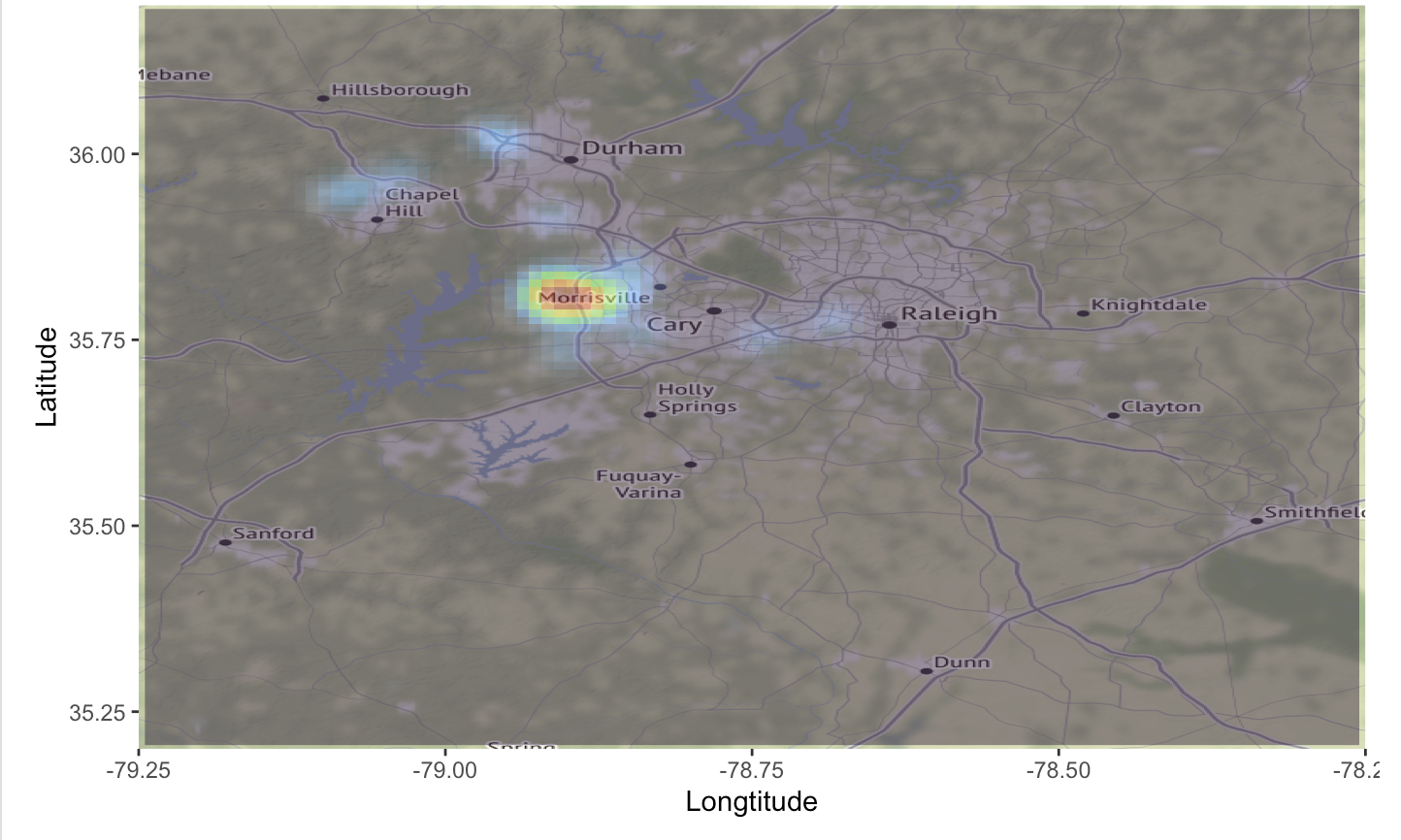}
    \captionof{figure}{Population density for respondents with duration $<$ 5 years.}
    \label{fig:duration1}
   \end{minipage}\hfill
   \begin{minipage}{0.48\textwidth}
     \centering
      \includegraphics[width = 8cm]{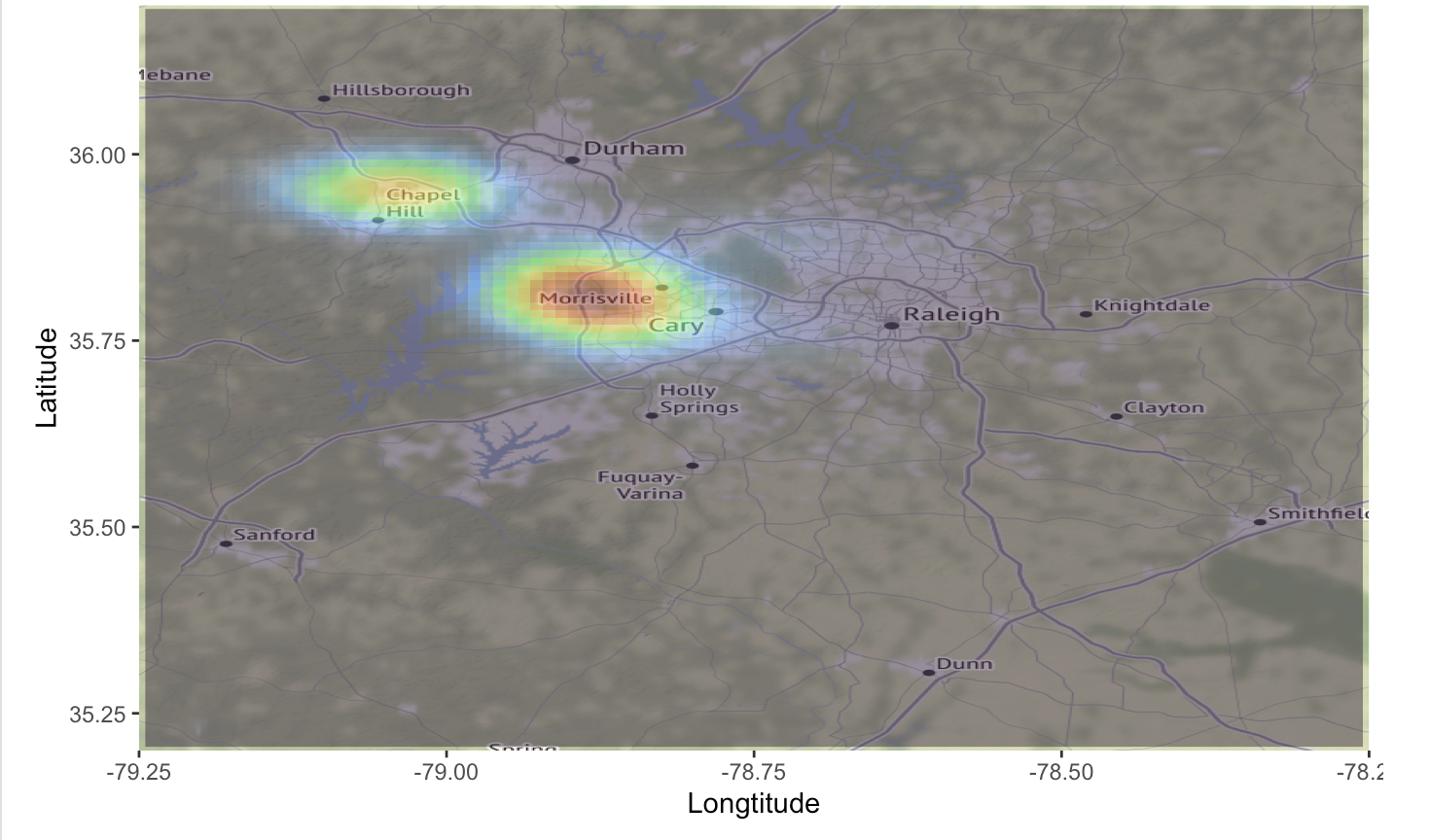}
    \captionof{figure}{Population density for respondents with duration $\geq$ 5 years.}
    \label{fig:duration2}
   \end{minipage}
 
\end{figure}

We next consider the geographic distribution of ChIRDU participants' residences by their duration, i.e., the total time in years the participant has lived in the Raleigh-Durham area. Duration of residence is a useful heuristic for specific social processes of immigrants such as identification and marriage \citep{treas}.
 For $j = 1, \dots, n_B$, let $\hat{Z}_j$ be the point estimate of $Z_j$, computed here as the value $i \in \{1, \dots, n_A, n_A+j\}$ giving the highest posterior probability.  Figure \ref{fig:duration1} and Figure \ref{fig:duration2} display the geographic densities of $\{\hat{Z}_j: j \leq n_A\}$, i.e., the residences for linked individuals, when duration is split at five years.   Respondents with duration less than five years exhibit a higher concentration in proximity to city centers, suggesting a preference for urban living or proximity to research centers/universities. 
 For respondents with longer duration (five or more years), we see higher population  density in suburban areas where home ownership is more affordable with more new, single-family, owner occupied housing, possibly indicating a tendency to establish more permanent residences and start or accommodate  families.  We note that the ChIRDU respondents with duration $< 5$ years are less likely to be married with children (63\%) and on average 7.5 years younger than those with duration exceeding five years (81\% married with children).

\begin{figure}[t]
\centering
\includegraphics[width=0.6\textwidth]{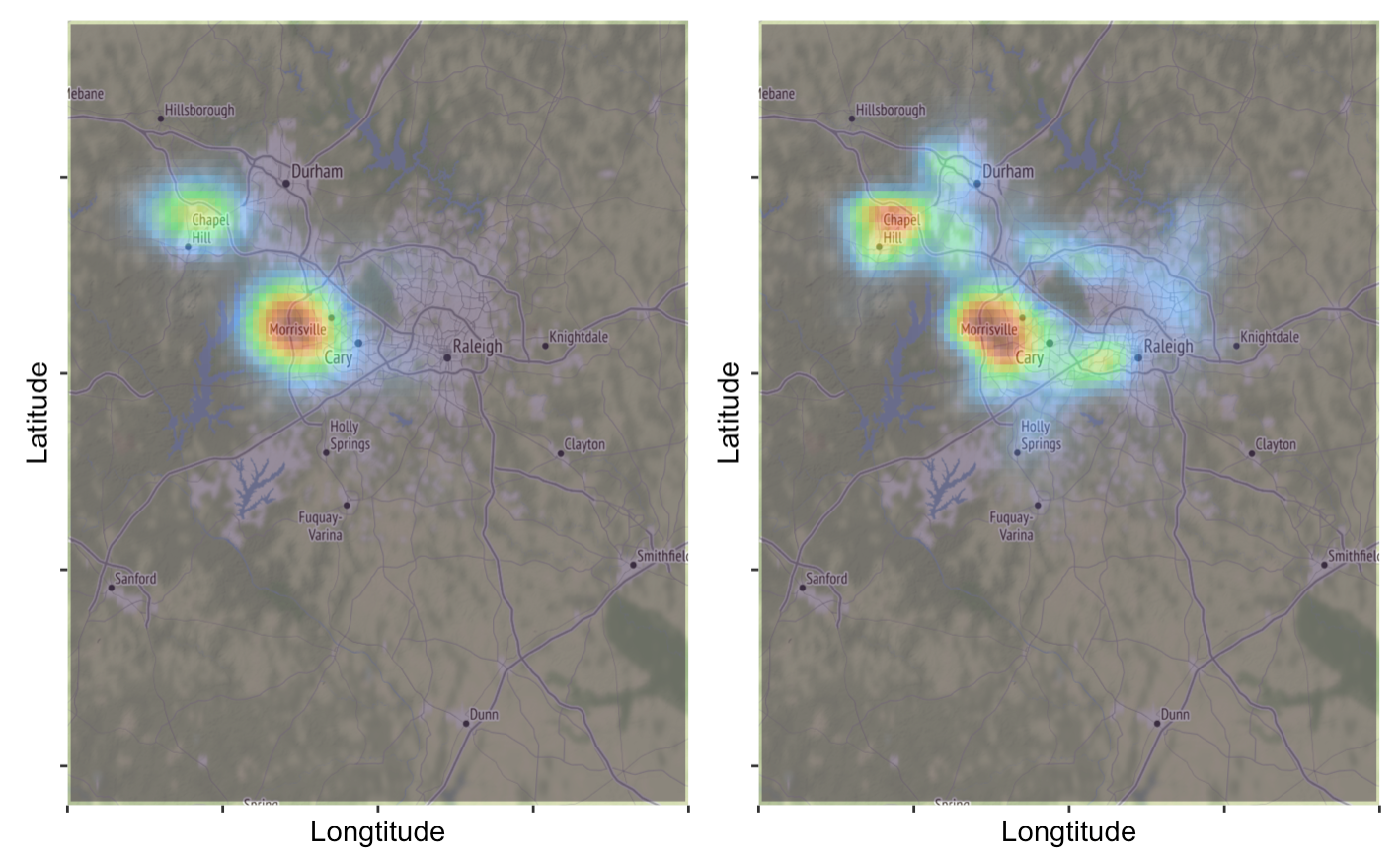}
\captionsetup{justification=centering}
\caption{Kernel density of linkages: posterior mode vs all eligible residences in InfoUSA.}
\label{fig:density_map_posteriorMode_infoUSAl}
\end{figure}

To assess spatial coverage of the linked sample, we compare the geographic distribution of the addresses of the linked ChiRDU participants to the spatial distribution of addresses in our subset of InfoUSA. 
 Figure \ref{fig:density_map_posteriorMode_infoUSAl} displays the density of ChIRDU  residences using   
 $\{\hat{Z}_j: j \leq n_A\}$, i.e., the most likely residences for linked individuals in ChIRDU, along with the density of residences in our InfoUSA sample.  Most individuals in the ChIRDU and InfoUSA data reside predominately in Morrisville/Cary and Chapel Hill.  Outside these areas, the linked ChIRDU cases are represented at somewhat lower frequency than in InfoUSA; however, with only 509 ChIRDU participants, it is not surprising to see low density outside the two main towns. Additionally, the ChIRDU data exclude Chinese international  students---since their immigration propensity is not known---who may appear in the InfoUSA data, particularly near university areas, but not among ChIRDU participants. This also may contribute to the relatively sparser spatial coverage among the linked ChIRDU records in proximity of Duke University and NC State University.  Nonetheless, we conclude overall that the spatial distribution among the linked ChIRDU records is similar enough to those in InfoUSA to learn about the characteristics of Chinese immigrants in the Raleigh-Durham study area or, using the InfoUSA ethnicity field, the characteristics of neighborhoods where Chinese immigrants reside.

The posterior draws of $\mathbf{Z}$ also facilitate assessment of the uncertainty in the linkages, including whether the model performs differentially by demographic groups, i.e., investigations of data equity within the linkage context. We provide these analyses in the supplementary material.


\section*{Concluding Remarks}\label{sec:conclusion}
 
 We present an approach to constructing comparison vectors to account for different transliterations in Chinese names. 
 We use these comparison vectors in a Bayesian record linkage model to match records from a  population-representative sample of Chinese immigrants  to records from a commercial database.  The resulting probabilistic record linkage yields a three-fold increase in likely matches compared to requiring exact matches on all four linking variables.  This enables us to add fine-resolution geographic detail to the ChIRDU data, which in turn facilitates more precise investigation of the spatial distributions of the Chinese immigrant population in the Raleigh-Durham area using analyses that would not be possible without the  linkages.

 
 Our approach to constructing comparison vectors can be applied in contexts beyond the ChIRDU and InfoUSA linkage.  Researchers seeking to link to records in commercial data may confront similar challenges, particularly for Chinese names, as commercial providers may not describe their processes of transliterations as part of metadata. If researchers can identify sets of name equivalents, they can construct four-level comparison vectors tailored to their own settings for any type of name transliteration.

The Bayesian model  provides estimates of uncertainty for the linkages, which in turn enables estimation of uncertainty in subsequent analyses that rely on the links. In particular, researchers can compute estimates of the quantities of interest using each plausibly linked data file, and combine the estimates using the inferential methods from multiple imputation for missing data \citep{rubin:1987}, as exemplified in e.g., \citet{dalzell:energy},  \citet{gutman:multiple}, and  \citet{guhajspi}.

\bibliography{report}   

\end{spacing}

\newpage

\begin{table}[t]
\caption{Illustration of the construction of the comparison vector for last names using artificial records.} 
\label{table:toy comp}
\begin{center} 
\scalebox{0.6}{
\begin{tabular}{|l|l|l|l|l|l|l|} 
\hline
\rule[-1ex]{0pt}{3.5ex}  \textbf{Index j} & \textbf{Index i} & \textbf{ChIRDU Surname in \emph{pinyin}} & \textbf{ChIRDU Surname in Wade-Giles} & \textbf{InfoUSA Surname} & \textbf{Jaro-Winkler} &\textbf{$\gamma^1_{i,j}$}\\
\hline\hline
\rowcolor{green!30}
\rule[-1ex]{0pt}{3.5ex}  1 & 1 & Wang && Wang &1.0 & 1 \\
\hline
\rule[-1ex]{0pt}{3.5ex}  1 & 2 & Wang && Tsai &0.5 & 4 \\
\hline
\rule[-1ex]{0pt}{3.5ex}  1 & 3 & Wang &&Chen &0.5 & 4 \\
\hline
\rule[-1ex]{0pt}{3.5ex}  1 & 4 & Wang && Zheng & 0.63& 4 \\
\hline
\rule[-1ex]{0pt}{3.5ex}  2 & 1 & Cai & Tsai &Wang &0.53& 4 \\\rowcolor{yellow!50}
\hline
\rule[-1ex]{0pt}{3.5ex}  2 & 2 & Cai & Tsai &Tsai &0.72& 2 \\ 
\hline
\rule[-1ex]{0pt}{3.5ex}  2 & 3 & Cai & Tsai &Chen &0.57 & 4 \\
\hline
\rule[-1ex]{0pt}{3.5ex}  2 & 4 & Cai & Tsai &Zheng & 0& 4 \\
\hline
\rule[-1ex]{0pt}{3.5ex}  3 & 1 & Zhen & Chen& Wang & 0.5& 4 \\
\hline
\rule[-1ex]{0pt}{3.5ex}  3 & 2 & Zhen & Chen &Tsai & 0.0 & 4 \\\rowcolor{yellow!50}
\hline
\rule[-1ex]{0pt}{3.5ex}  3 & 3 & Zhen & Chen & Chen &0.83& 2 \\\rowcolor{cyan!30}
\hline
\rule[-1ex]{0pt}{3.5ex}  3 & 4 & Zhen & Chen & Zheng & 0.96 & 3 \\
\hline
\end{tabular}
}
\end{center}
\end{table}

\clearpage
\appendix

\section{Supplemental Material for the Research Note: Bayesian Record Linkage with Application to Chinese Immigrants in Raleigh-Durham (ChIRDU) Study}

\begin{spacing}{2}   

\section*{Introduction}

In this supplement, we review techniques for probabilistic record linkage developed by \citet{fellegi1969theory}, which underpin the Bayesian record linkage model we use for the ChIRDU--InfoUSA linkage.  For thorough reviews of record linkage methods, we refer readers to \citet{binette_alm}, \citet{christian}, and \citet{scheurenwinkler}.  To keep this supplement self-contained, we repeat verbatim the fundamental setup described in the main text, followed by the descriptions of the record linkage models.  We also present the investigations of the linkage quality for different subgroups in the ChIRDU data.  Finally, we provide some key provisions from our approved IRB protocol.

\section*{Bayesian Record Linkage Model}
 Let $\mathcal{A}$ and $\mathcal{B}$ be two databases with $n_{A}$ and $n_{B}$ records, respectively.  Both $\mathcal{A}$ and $\mathcal{B}$ include $K$ variables in common, which we refer to as linking fields.  Let $\mathcal{A}= \{\mathcal{A}_i: i=1, \dots, n_A\}$, where  each record $\mathcal{A}_i = (\mathcal{A}_{i1}, \dots, \mathcal{A}_{iK})$ is measured on the $K$ variables.  Similarly, let $\mathcal{B}= \{\mathcal{B}_j: j=1, \dots, n_B\}$, where  each record $\mathcal{B}_j = (\mathcal{B}_{j1}, \dots, \mathcal{B}_{jK})$.   Our setting is a bipartite record linkage task, in which  
 each individual is represented at most once within each of $\mathcal{A}$ or $\mathcal{B}$, that is, no duplication exists within a database.  Both $\mathcal{A}$ and $\mathcal{B}$ may have other variables as well.  We do not concern ourselves with such variables for purposes of record linkage, although it is possible to utilize them to improve record linkage quality \citep[e.g., ][]{gutman, dalzellreiter, larsenlahiri, wortmanreiter}.



Intuitively, two records $\mathcal{A}_i$ and $\mathcal{B}_j$ that belong to the same individual should have similar values of most of the $K$ linking fields, whereas two records that belong to different individuals should not. 
We formalize this intuition by using comparison vectors.  
For every record $\mathcal{A}_i$ and $\mathcal{B}_j$,  and every linking field $k=1, \dots, K$,  
we define $\gamma_{ij}^k$ to take on levels of agreement in $\{1, \dots, d_k\}$. Here,  smaller values of $\gamma_{ij}^k$ indicate closer agreement on field $k$, and larger values indicate the opposite. For example, suppose we use zip code as the $k$th linking field, and we define the comparison as identical zip codes or not.  We would define a binary  $\gamma_{ij}^k$, i.e., with $d_k=2$.  We let $\gamma^k_{ij} = 1$ when records $\mathcal{A}_i$ and $\mathcal{B}_j$ have identical values of zip code,
and $\gamma^k_{ij} = 2$ otherwise. We define the collection of all comparison vectors as  $\gamma = \{\gamma_{ij}: i = 1, \dots, n_A; j = 1, \dots, n_B\}$.


For linking fields that involve strings, e.g., names or addresses, it is common to define the corresponding $\gamma_{ij}^k$ by using string similarity/distance metrics. 
This a function that maps two vectors of characters to a non-negative number, with higher values indicating greater similarity between the vectors. We then discretize these scores into distinct threshold/cutoffs to facilitate record linkage modeling. 
Several well-established string similarity metrics utilize edit-based distance, which counts the number of edit operations (substitution, deletion, insertion) required to transform  one string into another. The Jaro-Winkler metric, introduced by \citet{jaro} and improved by \citet{winkler}, gives heavier weight to the prefix similarities of strings and is more suitable on names where abbreviations/typos may occur at the beginning of the string. Another popular edit-distance metric is the Levenshtein distance, which calculates the minimum number of single-character edit operations required to transform one string into another \citep{levenshtein1966binary}. While most edit distance metrics have found success in modeling westernized words, we discuss their limitations in measuring romanized strings with non-western origins (romanized Chinese names) in the main text.  For a through survey on string matching, we refer readers to \citet{christian}.



\subsection*{Fellegi-Sunter Model for Probabilistic Record Linkage}
\label{sec:Frequentist FS}

We use $\gamma$ to decide which record pairs match and which do not.  For any record pair $(i,j)$, where $i \in \{1, \dots, n_{A}\}$ and $j \in \{1, \dots, n_{B}\}$, let 
 $C_{i,j}$ indicate whether $\mathcal{A}_i$ and $\mathcal{B}_j$ are a match. Specifically, 
\begin{equation} \label{c}
C_{ij}=
    \begin{cases}
        1, & \text{if record $\mathcal{A}_i$ matches record $\mathcal{B}_j$;} \\
        0, & \text{otherwise.} 
    \end{cases}
\end{equation}

In general, we do not know $C_{ij}$ for each record pair and need to estimate it from $\mathcal{A}$ and $\mathcal{B}$.
To do so, \citet{fellegi1969theory} propose a  mixture model, with one set of parameters for matches and another set of parameters for nonmatches.  The mixture model approach is widely used in probabilistic record linkage \citep[e.g., ][]{belinrubin, larsenrubin,enamorado, murray} including by national statistical agencies \citep{carra, nchs}.  

For any record pair $(i,j)$, let $\Gamma_{ij}=(\Gamma_{ij}^1, \dots, \Gamma_{ij}^K)$ be the vector of random variables for that record pair's $K$ comparison fields.  Thus, each $\gamma_{ij}^k$ is a realization of $\Gamma_{ij}^k$, and each $\gamma_{ij}$ is a realization of $\Gamma_{ij}$.  The mixture model specifies
\begin{align} 
&\Gamma_{ij}|C_{ij}=1 \sim G(\mathbf{m}) \label{generativem}\\
&  \Gamma_{ij}|C_{ij}=0 \sim G(\mathbf{u}). \label{generativeu}
\end{align}
Here,  $G(\mathbf{m})$ and $G(\mathbf{u})$ are the distributions  for the matches and nonmatches, respectively, with parameters $\mathbf{m} = (\mathbf{m}_1, \dots, \mathbf{m}_K)$ 
    and $\mathbf{u}=(\mathbf{u}_1, \dots, \mathbf{u}_K)$.   Each $\mathbf{m}_k = (m_{k,1}, \dots m_{k,d_k})$, where $m_{k,l} = p(\Gamma_{ij}^k=l | C_{ij}=1)$ is the probability that field $k$ takes on value $l$ for matches, for any $k$ and $l = 1, \dots, d_k$.  Similarly, each $\mathbf{u}_k = (u_{k,1}, \dots u_{k,d_k})$, where  $u_{k,l} = p(\Gamma_{ij}^k=l | C_{ij}=0)$ is the probability that field $k$ takes on value $l$ for nonmatches, for any $k$ and $l = 0, \dots, d_k$.


\citet{fellegi1969theory} assume the linking variables are conditionally independent, given matching status.   Thus,  we can expand \eqref{generativem} and \eqref{generativeu} to write the probabilities of observing $\gamma_{ij}$ as 
\begin{eqnarray} \label{eq:likelihood_sadinle}
    &  p(\Gamma_{ij}=\gamma_{ij} \mid C_{ij}=1) = \prod_{k=1}^{K}\prod_{l=0}^{d_k} m_{k,l}^{\mathbbm{1}(\gamma_{i,j}^k=l)}\\
    &  p(\Gamma_{ij}=\gamma_{ij} \mid C_{ij}=0)  =\prod_{k=1}^{K}\prod_{l=0}^{d_k} u_{k,l}^{\mathbbm{1}(\gamma_{i,j}^k=l)}.\label{eq:likelihood_sadinleU}
\end{eqnarray}
For convenience, we refer to the probabilities in \eqref{eq:likelihood_sadinle} and \eqref{eq:likelihood_sadinleU} as $\mathbf{m}(\gamma_{ij})$ and $\mathbf{u}(\gamma_{ij})$, respectively. 
Finally, \citet{fellegi1969theory} assume that each $\Gamma_{ij}$ is independent.  Thus, letting $\Gamma$ represent the collection of all pairs' $\Gamma_{ij}$,  their final model specifies
\begin{equation}
p(\Gamma = \gamma \mid \mathbf{m}, \mathbf{u}) = \prod_{i=1}^{n_A}\prod_{j=1}^{n_B}\mathbf{m}(\gamma_{ij})^{C_{ij}} \mathbf{u}(\gamma_{ij})^{1-C_{ij}}.\label{eq:finalFS}
\end{equation}
One consequence of the independence assumption for the $\Gamma_{ij}$ is that it is possible for some records in one of the files to match to multiple records in the other file. To make a one-to-one linkage, analysts can use optimization routines like those in \citet{jaro}.



\subsection*{Bayesian Versions of the Fellegi Sunter Model}
\label{sec:bayes}



The model of \citet{fellegi1969theory} results in a single point estimate of the linkage structure. In many cases, it can be beneficial for the point estimate to be accompanied by an uncertainty estimate \citep{reiteradmin}.  This can be done by using a Bayesian version of the  \citet{fellegi1969theory} model, as we do here.  In particular, we use the model presented by  \citet{sadinle}, which we  now  review.  For more information on Bayesian approaches to record linkage, we refer readers to \citet{fortini2001bayesian}, \citet{larsen2005hierarchical}, \citet{steortshall}, \cite{tangetal}, and \citet{guhacausal}, among others. 



\citet{sadinle} works with a different characterization of the linkage structure, which we also use.  Suppose that we seek to match records in $\mathcal{A}$ (in our context, InfoUSA) to records in $\mathcal{B}$ (in our context, ChIRDU), and that $n_A>n_B$. We define the co-referent vector $\mathbf{Z} = (Z_1, \dots, Z_{n_B})$ such that, for each $\mathcal{B}_j$ where $j=1, \dots, n_B$,  
\begin{align*}
Z_j=
    \begin{cases}
        i, & \text{if $\mathcal{B}_j$ matches to $\mathcal{A}_i$;} \\
        n_A+j, & \text{if $\mathcal{B}_j$ does not have a match in $\mathcal{A}$.} 
    \end{cases}
\end{align*}
The $\mathbf{Z}$ fully defines the linkage structure for all $n_An_B$ record pairs, since $C_{ij} = \mathbbm{I}(Z_j=i)$.  But, it is a much sparser representation, which facilitates efficient computation \citep{sadinle}.

Using $\mathbf{Z}$, we can write \eqref{eq:finalFS} as 
\begin{equation}
\label{eq:likelihood}
    p(\Gamma = \gamma \mid \mathbf{m}, \mathbf{u}) 
    = \prod_{i=1}^{n_A}\prod_{j=1}^{n_B}\prod_{k=1}^{K}\prod_{l=0}^{d_{k}-1}[m_{k,l}^{\mathbbm{1}(Z_j=i)}u_{k,l}^{\mathbbm{1}(Z_j \neq i)}]^{\mathbbm{1}(\gamma_{i,j}^k=l)}.
\end{equation}
We use \eqref{eq:likelihood} as the data model for the Bayesian record linkage model.


We also
require prior distributions on $\mathbf{m}, \mathbf{u},$ and $\mathbf{Z}$.  For 
each field $k$, the prior distributions for $\mathbf{m}_k$ and $\mathbf{u}_k$ are assumed to be independent Dirichlet distributions,
\begin{eqnarray} \label{eq:prior2m}
        \mathbf{m}_k &\sim& Dirichlet(\alpha_{k,1}, \dots, \alpha_{k,d_k}) \\
        \mathbf{u}_k &\sim& Dirichlet(\beta_{k,1}, \dots, \beta_{k,d_k}).\label{eq:prior2u}
\end{eqnarray}
Here, we set $\alpha_{k,l}=1$ and $\beta_{k,l}=1$ for all $k$ and $l$. This represents flat prior distributions on the $\mathbf{m}$ and $\mathbf{u}$ parameters.

For the prior distribution on $\mathbf{Z}$, we enforce the bipartite matching constraint using the beta prior for bipartite matchings developed by \citet{sadinle}.  
Specifically, he assumes independent Bernoulli distributions for the indicators of whether record $\mathcal{B}_j$ has a match, that is, $\mathbbm{1}(Z_j \leq n_A) \sim Bernoulli(\pi)$.  He also assumes that $\pi \sim Beta(\alpha_{\pi}, \beta_{\pi})$. 
Let the number of links be defined as $n_{AB}(\mathbf{Z}) = \sum_{j=1}^{n_B}\mathbbm{1}(Z_j \leq n_A)$.  Conditional on $n_{AB}(\mathbf{Z})$, all possible bipartite matchings are taken to be equally likely.  Putting it all together, \citet{sadinle} shows that 
this prior distribution corresponds to 
\begin{equation} \label{eq:prior}
    p(\mathbf{Z}|,\alpha_{\pi}, \beta_{\pi}) = \frac{(n_1-n_{\mathcal{AB}}(\mathbf{Z}))!}{n_\mathcal{A}!} \frac{B(n_{\mathcal{AB}}(\mathbf{Z}) + \alpha_{\pi}, n_\mathcal{B}-n_{\mathcal{AB}}(\mathbf{Z}) + \beta_{\pi})}{B(\alpha_{\pi}, \beta_{\pi})},
\end{equation}
where $B(\cdot,\cdot)$ is the Beta function. 
In the ChIRDU application, we set $\alpha_\pi =1$ and $\beta_\pi = 1$ to induce a uniform prior distribution.


With the likelihood in \eqref{eq:likelihood} and the prior distributions in (\ref{eq:prior2m}),  (\ref{eq:prior2u}), and (\ref{eq:prior}), we can use a Gibbs sampling algorithm \citep{sadinle} to generate many draws---we use $H=9000$ draws---from the posterior distribution of $\mathbf{Z}$. 
This posterior distribution facilitates uncertainty quantification on the linkages.  In particular, for any $\mathcal{B}_j$, we can compute the fraction of the $H$ draws where $Z_j=i$ as an estimate of the posterior probability that $\mathcal{A}_i$ matches to $\mathcal{B}_j$; we refer to this quantity as $p_j$.  We also can compute the fraction of times $Z_j=n_A+j$ to estimate the probability that $\mathcal{B}_j$ does not have a link in $\mathcal{A}.$  In applications where a point estimate is demanded, We can summarize $\mathbf{Z}$ via the posterior mode or Bayes estimates with suitable loss functions, with the latter taking on a simple closed form with additional assumptions \citep{sadinle}. For simplicity,  we summarize $\mathbf{Z}$ using the posterior mode whenever point estimates are required for the analysis.  

Finally, we note that the model of \citet{sadinle} can be estimated using the ``BRL'' package in R, which we do in the ChIRDU and InfoUSA linkage.

\section*{Evaluation of Linkage Quality}\label{sec:qual}

 The draws of $\mathbf{Z}$ also allow us to examine the quality of the linkages.  We begin by evaluating linkage rates  across various demographic groups.
The goal is to to uncover any disparities in matching outcomes across groups. 
We characterize the quality of linkages by their average posterior link probabilities, where higher probabilities imply greater faith in the linkages. 

For any demographic group $g$, let $\mathcal{G}$ include the $n_g$ ChIRDU participants in that group.   
We compute
\begin{equation}
{p}_{g} = \frac{1}{n_g}\sum_{j \in \mathcal{G}}\mathbbm{1}(\hat{Z}_j \leq n_A).
\end{equation}
This is the percentage of records in group $\mathcal{G}$ that the model deems likely to have a match in InfoUSA.

Table \ref{tab:duration} through Table \ref{tab:sex} display $p_{g}$ for demographic groups defined by duration in the Raleigh-Durham area, 
education level, age and sex.  
It is most difficult to find matches for individuals who have resided in the study area for less than 10 years, or who have at most a high school degree, or who are in the youngest or oldest age categories.  This may result from variation in the  quality of the available names.  For example, when collecting data, the ChIRDU investigators found that ChIRDU respondents 
with less than high school degree were more reluctant to provide full names, as well as to provide information  that might be used to infer incomplete names (e.g., email addresses).  
Nonetheless, for groups with substantial sample sizes, the match rates are reasonably close to the 40\% overall match rate.  



\begin{figure}[t]
\centering
\begin{minipage}[t]{0.45\linewidth}
  \centering
  \begin{table}[H]
    \centering
    \caption{Posterior Match Rates: Duration.}
    \label{tab:duration}
    \small 
    \setlength{\tabcolsep}{4pt} 
    \begin{tabular}{@{}cp{2.3cm}p{2.5cm}@{}} 
    \hline
    \textbf{Duration (Years) } & \textbf{Average Link Probability} & \textbf{\% of ChIRDU}  \\
    \hline
    0-10  & 0.30 & 0.64  \\
    10-20 & 0.57 & 0.26  \\
    20-30  & 0.49 & 0.09  \\
    30-40 & 0.37 & 0.01  \\
    \hline
    \end{tabular}
  \end{table}
\end{minipage}
\hfill
\begin{minipage}[t]{0.45\linewidth}
  \centering
    \begin{table}[H]
    \centering
    \caption{Posterior Match Rates: Education.}
    \label{tab:education}
    \small 
    \setlength{\tabcolsep}{4pt} 
    \begin{tabular}{@{}cp{2.3cm}p{2.5cm}@{}} 
    \hline
    \textbf{Education} & \textbf{Average Link Probability} & \textbf{\% of ChIRDU}  \\
    \hline
    High School or Less & 0.17 & 0.12  \\
    1-2 Years of College & 0.27 & 0.08  \\
    4 Year College & 0.33 &0.21  \\
    Postgraduate & 0.47 & 0.59  \\
    
    \hline
    \end{tabular}
  \end{table}
\end{minipage}
\vspace{0.5cm}

\begin{minipage}[t]{0.45\linewidth}
  \centering
    \begin{table}[H]
    \centering
    \caption{Posterior Match Rates: Age.}
    \label{tab:age}
    \small 
    \setlength{\tabcolsep}{4pt} 
    \begin{tabular}{@{}cp{2.3cm}p{2.5cm}@{}} 
    \hline
    \textbf{Age} & \textbf{Average Link Probability} & \textbf{\% of ChIRDU}  \\
    \hline
    Under 30 & 0.23 & 0.12  \\
    30-39 & 0.30 & 0.22  \\
    40-49 & 0.46 &0.35  \\
    50-59 & 0.47 & 0.22  \\
    60-69 & 0.40 & 0.06  \\
    70+ & 0.13 & 0.02  \\
    \hline
    \end{tabular}
  \end{table}
\end{minipage}
\hfill
\begin{minipage}[t]{0.45\linewidth}
  \centering
    \begin{table}[H]
    \centering
    \caption{Posterior Match Rates: Sex.}
    \label{tab:sex}
    \small 
    \setlength{\tabcolsep}{4pt} 
    \begin{tabular}{@{}cp{2.3cm}p{2.5cm}@{}} 
    \hline
    \textbf{Sex} & \textbf{Average Link Probability} & \textbf{\% of ChIRDU}  \\
    \hline
    Male & 0.43 & 0.35  \\
    Female & 0.37 & 0.65  \\
    \hline
    \end{tabular}
  \end{table}
\end{minipage}
\end{figure}

\begin{figure}[t]
\centering
\includegraphics[width=0.5\textwidth]{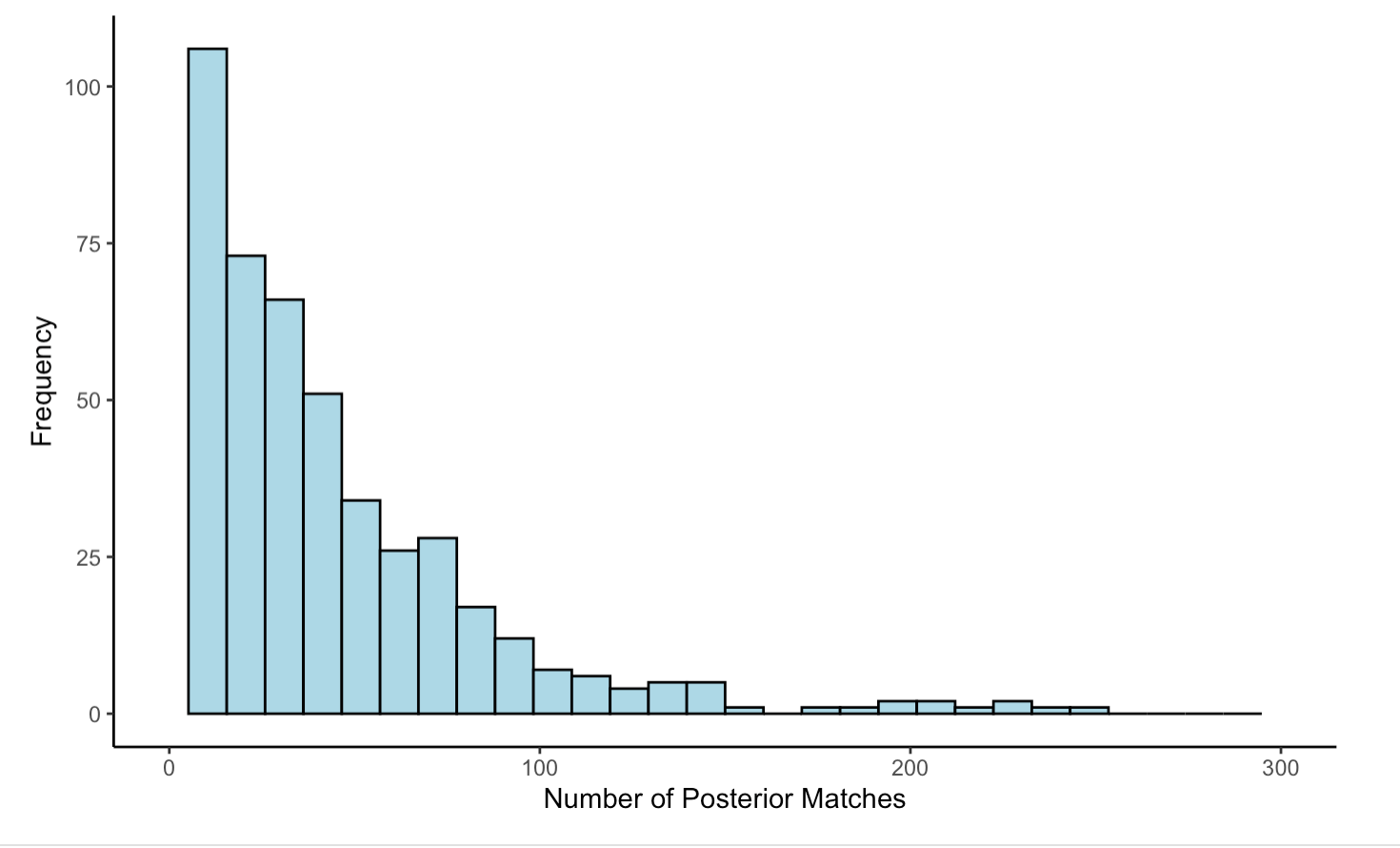}
\caption{Distribution of distinct matches.}
\label{fig:posterior_matches}
\end{figure}

We next examine the distribution of the number of distinct matches for each individual across the 9,000 iterations of the MCMC sampler. Higher uncertainty in the linkages is represented by high frequencies in the number of distinct posterior matches. Figure \ref{fig:posterior_matches} displays this distribution for the 509 individuals in ChIRDU. Of the 509 participants, only 93 ($\sim$18\%) of them have less than or equal to 10 distinct matches, and 191 ($\sim$38\%) have less than or equal to 20 distinct matches. In all, this suggests that there is substantial uncertainty in the linkages.

Counting the number of distinct matches without taking into account the relative frequency of each match can be misleading. For example, consider two hypothetical individuals Eric and Botao, where the former has five distinct matches and the latter has more than 100 distinct matches. Suppose each of the five matches for Eric has posterior probability of 20\%, whereas the top three matches for Botao have posterior probability of 70\%, 20\%, and 5\%. Though Botao displays more variability in the number of distinct matches, the relative frequency for his matches is less variable than that for Eric. 

To account for these issues, we develop a metric called the concentration ratio. 
For each $\mathcal{B}_j$, let $P_{Z_j} = \{p(Z_j = i| \gamma_{ij}): i\leq n_A\}$ be the set of all posterior probabilities for the matching records, ordered from largest to smallest value. We refer to the largest probability as $P_{Z_j}[1]$, the second largest as $P_{Z_j}[2]$, and so on. 
We define the concentration ratio for $\mathcal{B}_j$ at level $c$ as 
\begin{equation}
R^c_j = \sum_{k=1}^c P_{Z_j}[k].
\end{equation}
When $R^c_j$ is near 1 for small $c$, say $c\leq3$, the distribution of matches for $\mathcal{B}_j$ is concentrated around a small number of records in $\mathcal{A}$, suggesting we have good confidence that the record has a match in $\mathcal{A}$.  When $R^c_j$ is not near 1 when $c$ is small, we do not have good confidence that the record has a match.

\begin{figure}[t]
\noindent\begin{minipage}{0.48\textwidth}
     \centering
     \includegraphics[width = 8cm]{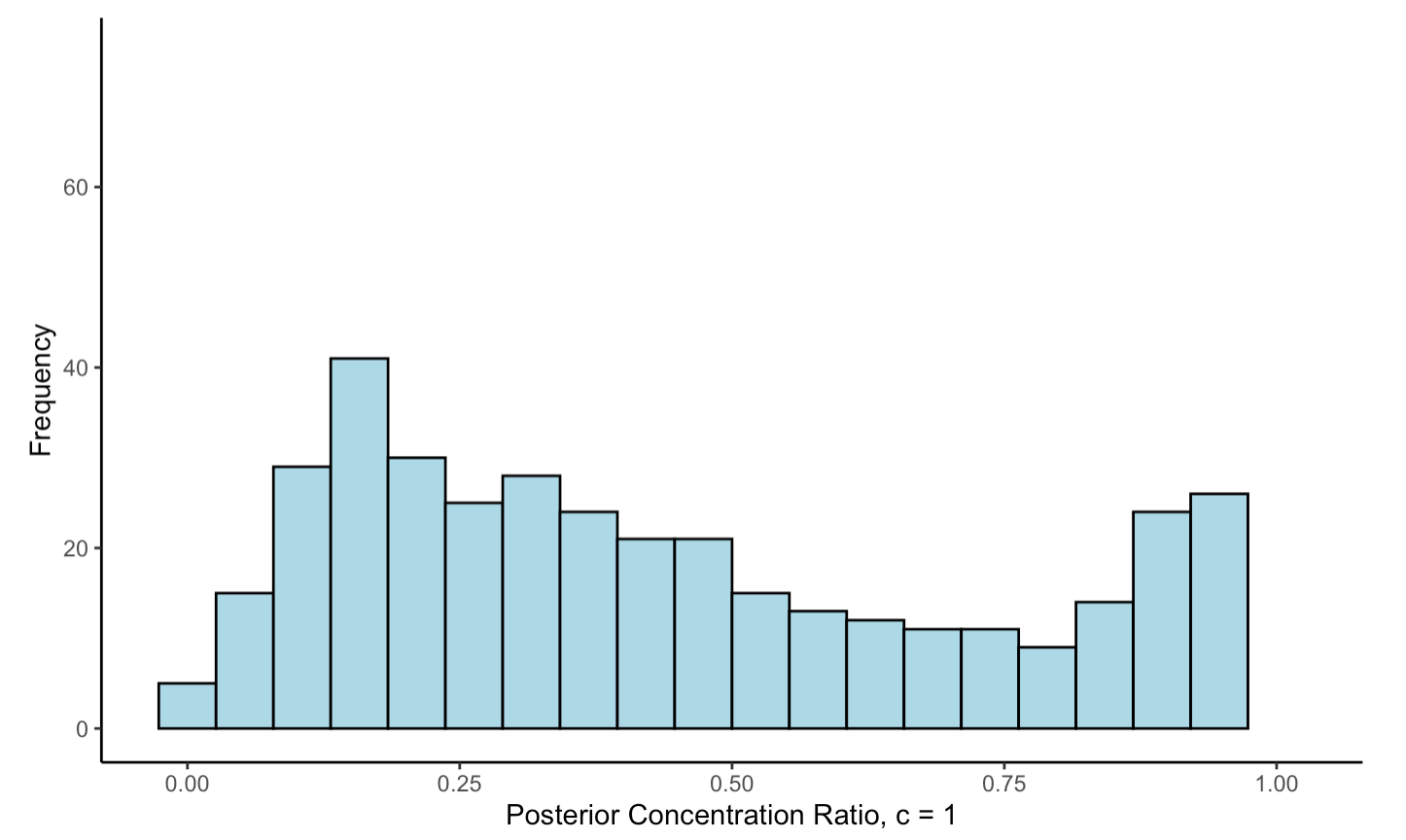}
    \captionof{figure}{Distribution of $R_1(j)$ for $c=1$.}
    \label{fig:CR1}
   \end{minipage}\hfill
   \begin{minipage}{0.48\textwidth}
       \centering
       \includegraphics[width = 8cm]{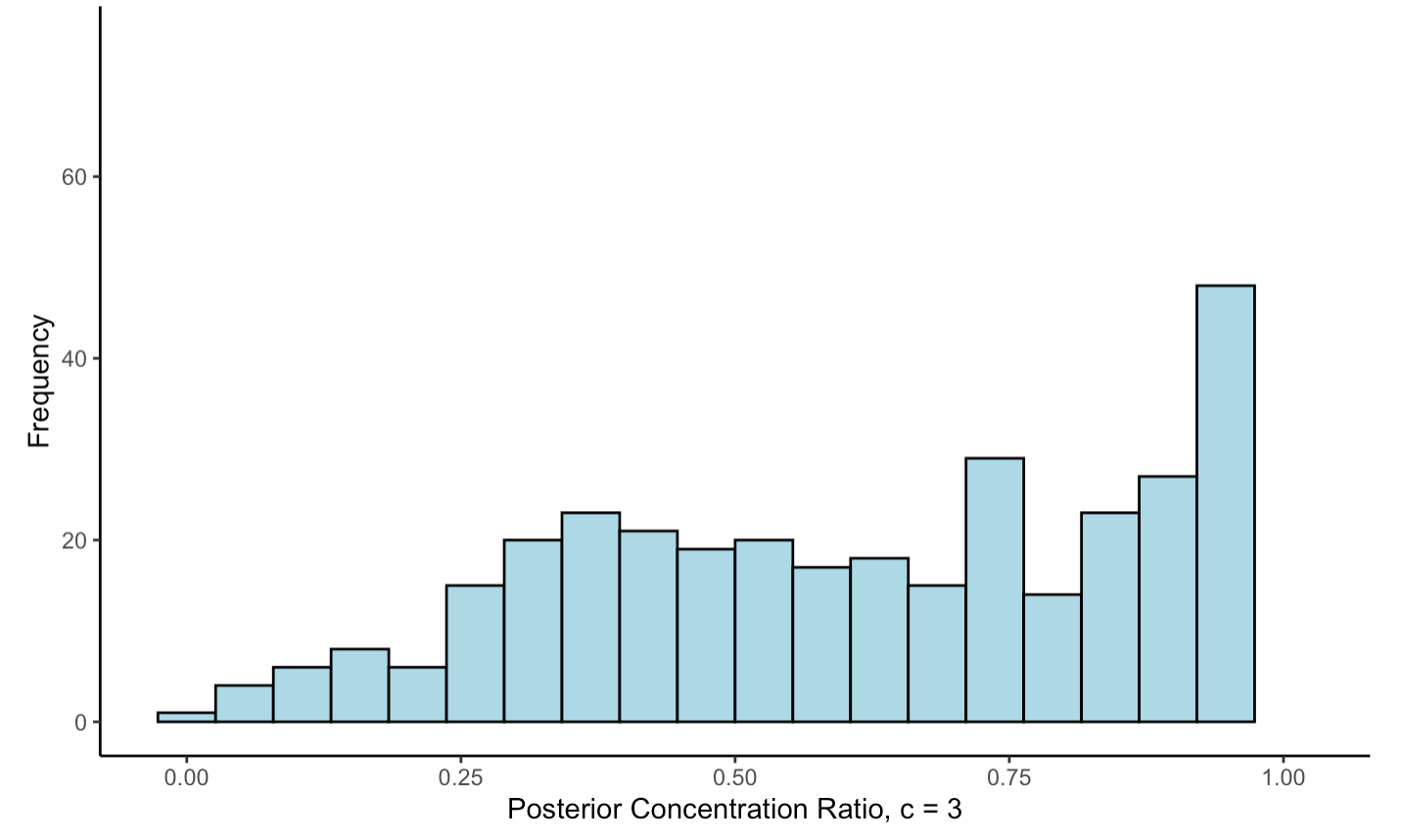}
       \captionof{figure}{Distribution of $R_3(j)$ for $c=3$.}
       \label{fig:CR3}
   \end{minipage}\hfill
\end{figure}

Figure \ref{fig:CR1} and Figure \ref{fig:CR3} display the concentration ratios for all ChIRDU participants when $c=1$ and $c=3$.  When $c=1$, the concentration ratios indicate that many records do not have a single match with high probability, e.g., above 50\%.  When $c=3$, the concentration ratios suggest that most records have at least 50\% probability of matching among the three highest links.  In all, these results once again reveal that there is substantial uncertainty in the record linkages, but that we have reasonable confidence that most records from ChIRDU  are in the InfoUSA data. 

\section*{Key Features of the IRB Protocol to Enable Linkages}

The following is taken directly from the approved IRB protocol for linking the ChIRDU and InfoUSA data.  It summarizes the steps to reduce disclosure risks during the linkage and data analysis process.  

``The analysis will proceed in several stages: 

(a) The research will start with a construction of a file from the ChIRDU data and a file from the InfoUSA data that only include the variables needed for probabilistic matching (names, gender, age, marital status, zip code, phone number). These are called the matching files. All other variables (including sensitive variables such as income and citizenship) will be excluded from the matching files. The matching files have direct identifiers—otherwise, the information for matching would be too diffuse to be useful—but no other information, which means that the matched records exclude sensitive variables such as income and citizenship. 

(b) Using the matching files, records from InfoUSA that have zero chance of being links to ChIRDU cases will be filtered out. These include individuals who do not live in the study region and individuals whose last names obviously cannot be matches for the names in our ChIRDU sample. This pre-processing step helps with computational efficiency and reduces risks for out-of-sample individuals. 

(c) Probabilistic record linkage techniques on the pre-processed matching files will be used to assess which records are (likely) matches across the two files. Once it is determined which records are matching across the two files, relevant study variables from the ChIRDU and InfoUSA files will be added back to the linked records and an analysis dataset created from which the direct identifiers (e.g., names) are removed and pseudo-IDs added, based on the protocols described in the approved ChIRDU protocol. Only this de-identified dataset will be used for analyses. After finalizing the analysis dataset, the matching files will be deleted so that only the de-identified analysis file can be stored and accessed for the analysis of linked data. This implies that the analysis dataset, which includes sensitive variables such as income and citizenship, will be de-identified. 

(d) To mitigate the possibility that an inadvertent release of the de-identified linked data could result in income and citizenship information for this sample of the RDU Chinese community being revealed, a record-level dataset will not be released to the public. Thus, the risks of inadvertent release of sensitive data beyond those in the current ChIRDU protocol should be minimal. 

(e) The data will be processed, stored and analyzed on secure [blanked for anonymity] systems. The two graduate students, Becker and Xu, 
will have access to the matching file in (a) but not any other information from either data source. The PI Reiter will not have access to the matching files with the direct identifiers in (a). Reiter will have access to the de-identified, analysis file created in (c). The graduate students will not have access to the analysis file created in (c). Merli, as co-investigator, will have temporary access to the identifiable linked dataset in (a) until the de-identified analysis dataset in (c) is created.''


\subsection* {Acknowledgments}
The ChIRDU data described here were collected with funds from NICHD grant R21HD086738 (Merli and Mouw, MPIs). The analyses are supported by pilot funds from NICHD grant 2P2CHD065563 to the Duke Population Research Center.

\end{spacing}

\end{document}